\documentstyle[aps,prl,psfig,multicol]{revtex}
\begin{document}
\draft

\title{Harmonic forcing of an extended oscillatory system: \\
Homogeneous and periodic solutions}

\author{Jeenu Kim,$^1$ Jysoo Lee,$^{2,3}$ and Byungnam Kahng$^1$}
\address{$^{1}$School of Physics and Center for Theoretical Physics,
Seoul National University, Seoul 151-742, Korea\\
$^{2}$National Creative Research Initiative Center for Neurodynamics
and Department of Physics,\\
Korea University, Seoul 136-701, Korea\\
$^{3}$Supercomputing Research Department, 
Korea Institute of Science and Technology Information,\\
P.O.Box 122, Yusong, Daejon 305-806, Korea}
\date{\today}
\maketitle

\begin{abstract}
In this paper we study the effect of external harmonic forcing on a
one-dimensional oscillatory system described by the complex Ginzburg-Landau
equation (CGLE).  For a sufficiently large forcing amplitude, a homogeneous
state with no spatial structure is observed.  The state becomes unstable
to a spatially periodic ``stripe'' state via a supercritical bifurcation
as the forcing amplitude decreases.  An approximate phase equation is derived,
and an analytic solution for the stripe state is obtained, through which
the asymmetric behavior of the stability border of the state is explained.
The phase equation, in particular the analytic solution, is found to be
very useful in understanding the stability borders of the homogeneous
and stripe states of the forced CGLE.
\end{abstract}


\begin{multicols}{2}

\section{Introduction}


Nonequilibrium pattern formation is widely observed in many physical,
chemical and biological systems.  Significant progresses have been
made in this field during the last few decades.  For example, it has
been found that nonequilibrium patterns can be grouped into a few
universality classes \cite{Cross,m90,Walgraef}.  In many cases, such a
system is in constant interaction with its environment, and
understanding the effect of extrinsic perturbation is of great
theoretical and practical importance.  In particular, it is
interesting to study the deformation of an existing pattern or the
formation of a new pattern under an external forcing.  Our
understanding in such a direction is far from complete.


Petrov {\em et al.}, and later Lin {\em et al.}, studied the light
sensitive Belousov-Zhabotinsky (BZ) reaction in an oscillatory regime
in the presence of a periodic modulation of the intensity of
illumination \cite{Petrov,Lin}.  They observed ``entrainment bands''
in which the system is frequency locked.  Different spatial
patterns---stationary fronts, standing waves of labyrinth and more
complex shapes---are observed within the bands.  In a similar BZ
reaction setup, Vanag {\em et al.}  studied spatial patterns and
transitions among them in detail, and observed localized
irregular/standing clusters as well as the above patterns
\cite{Vanag,Vanag2}.


Continuum models of forced pattern forming systems can be grouped into
ones based on a kinetic model or on an amplitude equation.  In the
first group, an unforced system is modeled by a coupled kinetic model,
such as the Brusselator, Oregonator or FitzHugh-Nagumo model, and
parameters in the model are modulated to simulate the effect of
external forcing (e.g. \cite{Lin,Steinbock}).  On the other hand, near
a bifurcation onset of a pattern, small differences among systems
become irrelevant, and they are all described by one of a few
universal equations.  If the bifurcation is supercritical and
oscillatory, and if the most unstable wave number is zero, the complex
Ginzburg-Landau equation (CGLE) is the equation for the class of
systems.  In the presence of an external periodic modulation, it is
shown that the CGLE with an additional forcing term becomes the
appropriate equation \cite{Walgraef,Coullet}.  There exist a few
studies on forced CGLE, and diverse behaviors are observed depending
on several factors, such as the spatial dimension, the mode of the
frequency locking, and the behavior of the corresponding unforced
system
\cite{Walgraef,Coullet,Lega,Elphick,Elphick2,Chate3,Lin3,Park,Gallego}.
However, we do not even know what behaviors are possible, let alone
understand them.

Even the simplest case of the 1:1 locking in one dimension displays a
wide variety of behaviors.  At large amplitude of the forcing, a
homogeneous state with no spatial structure is stable.  Chat{\'e} {\em
et al.} found that the homogeneous state becomes unstable to a
periodic ``stripe'' or ``kink-breeding'' state as the forcing
amplitude decreases, and a ``turbulent synchronized'' state---chaotic
with its average phase is locked to that of the forcing---can appear,
as the amplitude decreases further \cite{Chate3}.


In this paper, we study in detail the homogeneous and stripe states of
one dimensional forced CGLE around the 1:1 locking.  There are two
borders regarding the homogeneous state (1) the stability border,
below which the state loses its stability, and (2) the existence
border, below which a homogeneous solution does not exist.  In
general, the existence and stability borders do not coincide.  It is
known that the stability border of the homogeneous state lacks a
reflection symmetry around the $\nu = \alpha$ line.  Here, $\nu$ is
the difference between the natural and external frequencies, and
$\alpha$ is a nonlinear detuning parameter.  We find the asymmetry
can be explained by the linear stability of the state.  Also, the
condition under which the existence and stability borders of the
homogeneous state coincide is found.  The stability border of the
stripe state also lacks a reflection symmetry.  An approximate phase
equation is derived from the forced CGLE, and it is found that its
qualitative behavior is identical to that of the original equation, at
least in the region of present interest.  An analytic expression of
the stripe state for the phase equation is obtained, which is used to
explain the asymmetry of the border of the stripe state.


\section{Forced Complex Ginzburg-Landau Equation}

\subsection{Complex Ginzburg-Landau equation}

Near the stability border of a homogeneous state of an extended
pattern forming system, the time evolution in a large spatial and
temporal scale is given by one of a few universal equations
\cite{Cross,m90,Walgraef}.  If the instability is oscillatory and
supercritical, and the wave number of the most unstable mode is zero,
the CGLE,
\begin{equation}
\label{Eq:CGLE}
\partial_{t} A = A - (1 + i \alpha) |A|^2 A + (1 + i \beta)
\nabla^2 A
\end{equation}
is the governing equation.  Here, $A$ is complex amplitude, and
$\alpha, \beta$ are real constants.  The behavior of the CGLE is
relatively well understood, especially in one and two dimensions
\cite{Shraiman,Chate1,Chate2}.  It has plane wave solutions, which are
stable only if $1 + \alpha \beta > 0$.  Otherwise, the Benjamin-Feir
instability sets in, making the solutions unstable.  Near the unstable
side of the stability border ($1 + \alpha \beta = 0$ line), ``phase
turbulence'' is observed, which is characterized by disordered
cellular structure and the absence of a defect ($|A| = 0$).  ``Defect
turbulence'' is observed further in the unstable region, where
constant creation and annihilation of defects is observed
\cite{cl88,cgl89}.  In this paper, the value of $\alpha = -3/4, \beta
= 2$ is mainly used, which is in the phase turbulence region.

\subsection{Homogeneous state}

Consider the case wherein a sinusoidal forcing is applied to the system
of Eq.~(\ref{Eq:CGLE}).  It was shown that an additional forcing term
should be included, and its form can be determined from the conditions
of the spatial and temporal translation invariance \cite{Coullet}.
For a harmonic forcing (near the 1:1 tongue), the resulting equation
is
\begin{equation}
\label{Eq:FCGLE}
\partial_{t} A = (1 + i \nu)A - (1 + i \alpha) \left|A\right|^2 A
+(1+ i \beta) \nabla^2 A + B,
\end{equation}
where $\nu$ is the difference between the natural and forcing
frequencies, and $B$ is related to the amplitude of the forcing.

\begin{figure}[ht]
\centerline{\psfig{figure=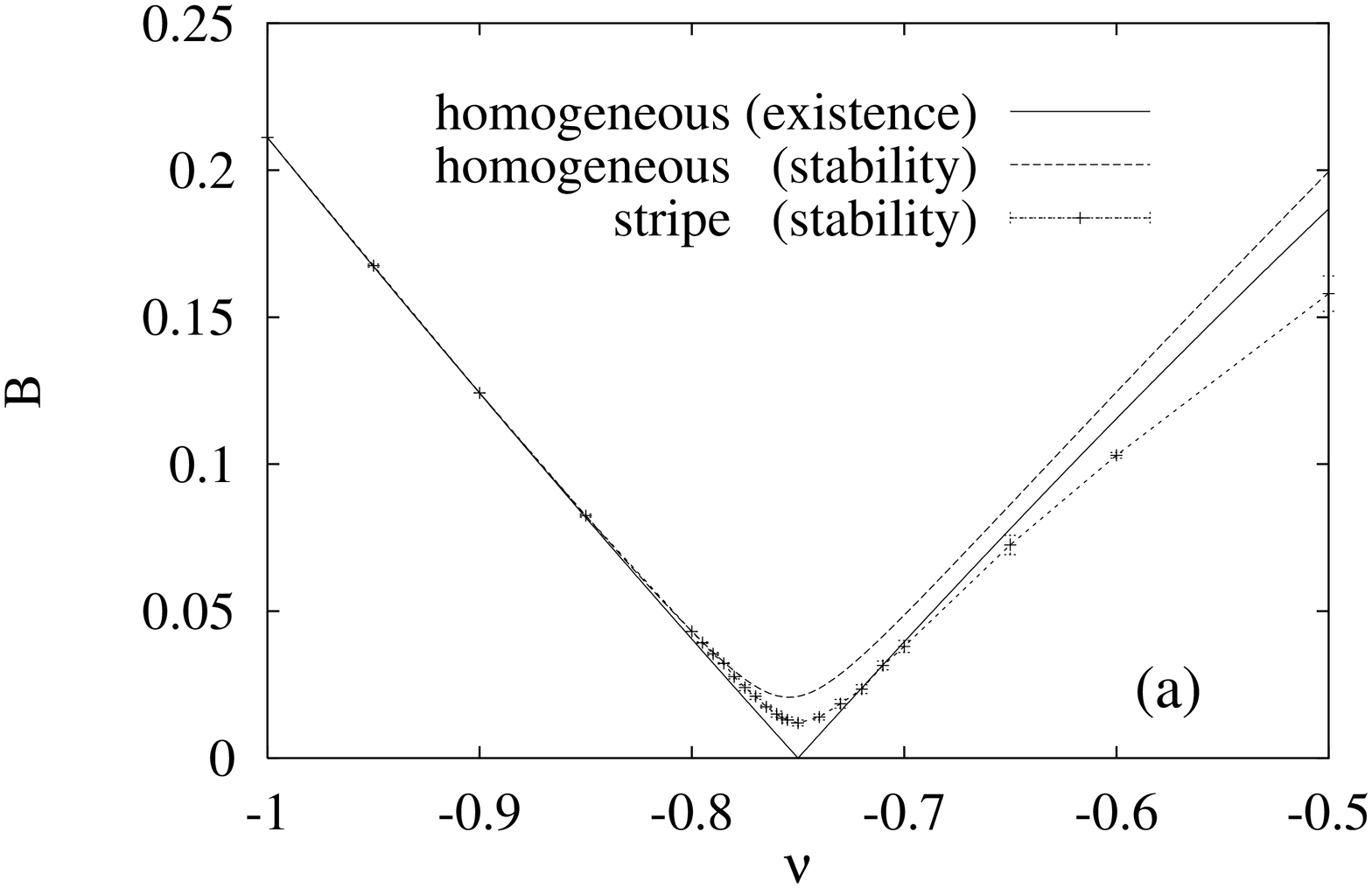,angle=270,width=0.45\textwidth}}
\centerline{\psfig{figure=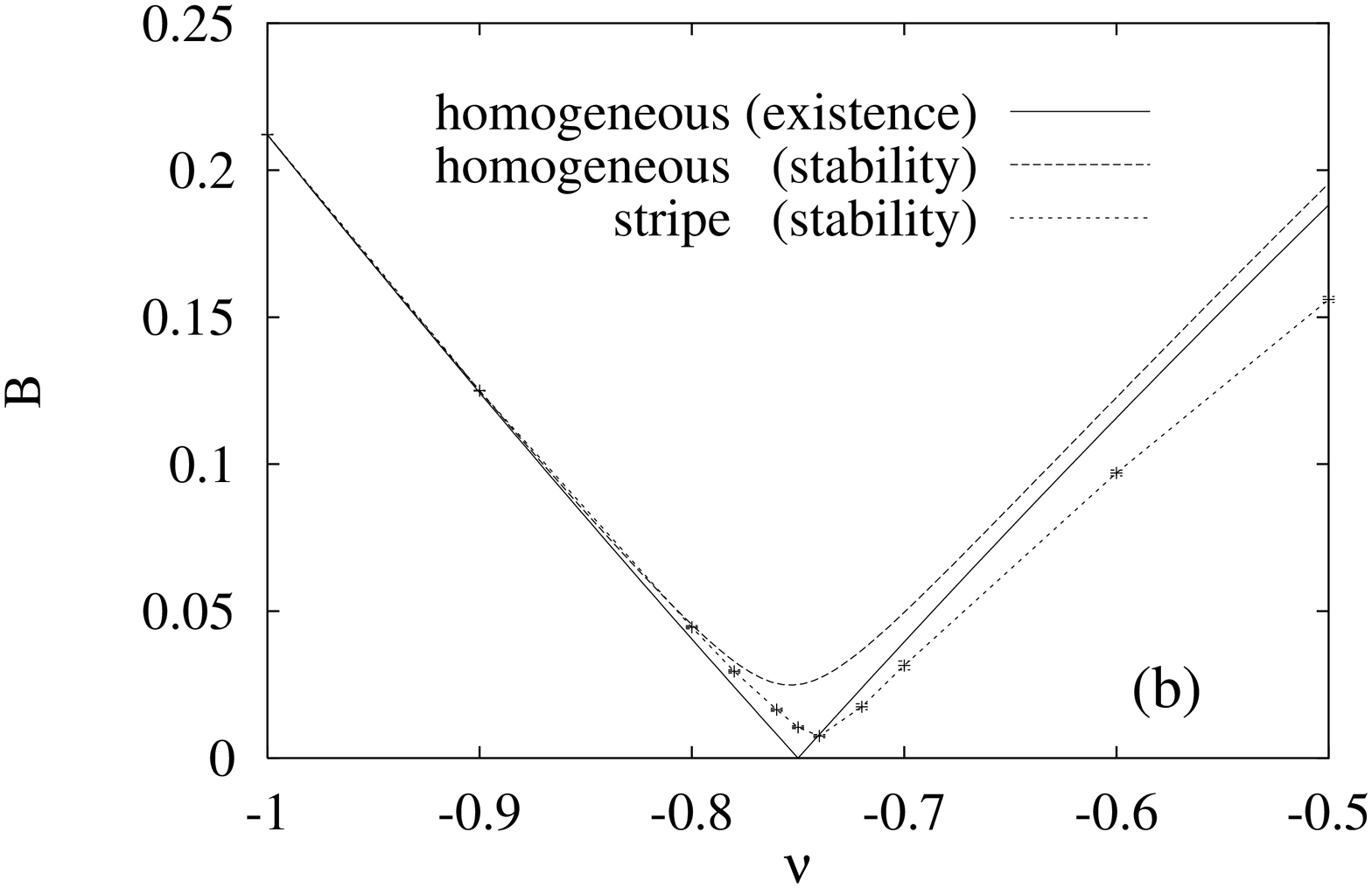,angle=270,width=0.45\textwidth}}
\caption{(a) The existence and stability borders of the homogeneous
state for the harmonically forced CGLE Eq.~(\ref{Eq:FCGLE}) with
$\alpha = -3/4$ and $\beta = 2$.  The stability border of the stripe
state is also shown.  Note that the stability borders are not
symmetric to the $\nu = \alpha$ line. (b) Corresponding borders for
the phase equation Eq.~(\ref{Eq:Phase}) with $\alpha = -3/4$, $\beta =
2$, and $R_0 = 1$.}
\label{fig:phase-d}
\end{figure}

We first seek for the homogeneous solution of Eq.~(\ref{Eq:FCGLE}).
In polar coordinates [$A = R \exp(i \Phi)$], the equation becomes
\begin{eqnarray}
\label{Eq:RP}
\partial_{t} R & = & R - R^3 + B \cos{\Phi} \nonumber\\ && + R_{xx} - \beta R
{\Phi}_{xx} - 2 \beta R_{x} {\Phi}_{x} - R{\Phi}_{
x}^2, \nonumber \\
R \partial_{t} \Phi & = & \nu R - \alpha R^3 -
B \sin{\Phi} \nonumber\\ && + \beta R_{xx} + R {\Phi}_{xx} + 2 R_{x}
{\Phi}_{x} - \beta R {\Phi}_{x}^2.
\end{eqnarray}
For a sufficiently large $B$, the system is expected to lock to the
forcing.  Its homogeneous solution is
\begin{eqnarray}
\label{Eq:R0}
B \cos{\Phi}_0 & = & - R_0 (1 - R_0^2), \nonumber \\
B \sin{\Phi}_0 & = & R_0 (\nu - \alpha R_0^2),
\end{eqnarray}
which can have one or three roots depending on the parameters.  For the
three root case, only the one corresponding to the largest $R_{0}$ is
stable.  The region of the parameter space in which a locked
homogeneous solution exists is shown in Fig.~\ref{fig:phase-d}(a).


We apply the linear stability analysis to the homogeneous solution
\cite{Chate3}, where the behavior of a small deviation from the
solution $r = R - R_{\rm 0}$ and $\phi = \Phi - \Phi_{\rm 0}$ is
studied.  The growth rate of the mode with wave number $k$ is found to
be
\begin{eqnarray}
\label{Eq:Rate}
\lambda(k) &=& 1 -2 R_0^2 -k^2 \nonumber\\
&&+ \sqrt {(1+\alpha^2)R_0^4 - (\nu -2\alpha R_0^2 -\beta k^2)^2},
\end{eqnarray}
which has the maximum value of
\begin{equation}
\label{Eq:Ratemax}
\lambda_{\rm max} = 1 -2 R_0^2 -{1 \over \beta} [\nu -\{ 2
\alpha +\sqrt{(1+\alpha^2)(1+\beta^2)} \} R_0^2 ]
\end{equation}
at $k = k_{\rm max}$, corresponding to the most unstable mode, which
is given by
\begin{equation}
\label{Eq:kmax}
\beta k_{\rm max}^2 = \nu -\left( 2\alpha +
\sqrt{\frac{1+\alpha^2}{1+\beta^2}} \right) R_0^2.
\end{equation}
The stability border $B_{\rm s}$ of the homogeneous solution is
obtained by solving numerically $\lambda_{\rm max} = 0$, which is also
shown in Fig.~\ref{fig:phase-d}(a).


A distinct feature of the stability border is that it is not symmetric
to the $\nu = \alpha$ line.  As shown in the figure, the difference
between the existence and the stability border is smaller at the $\nu <
\alpha$ side.  Moreover, the difference vanishes for $\nu \le \nu_{\rm
c}$ with $\nu_{\rm c} \simeq -1.067$.  This feature can be understood
from the $\nu$ dependence of $k_{\rm max}$, which is given by
Eq.~(\ref{Eq:kmax}).  It is found that $k_{\rm max}$ is an increasing
function of $\nu$: it is zero for $\nu \le \nu_{\rm c} =
[2\alpha+\sqrt{(1+\alpha^2)/(1+\beta)}]R_0^2$, and proportional to
$\sqrt{\nu - \nu_{\rm c}}$ slightly above $\nu_{\rm c}$.  Since the
wave number of the most unstable mode is zero for $\nu \le \nu_{\rm
c}$, and since the solution Eq.~(\ref{Eq:R0}) with the largest $R_0$
is stable to a zero wave number perturbation, the existence of the
homogeneous solution guarantees its stability.

As will be discussed later, an approximate phase equation is derived
from Eq.~(\ref{Eq:FCGLE}), which gives an additional insight into the
stability border.  The origin of the instability of the homogeneous
state of the phase equation can be traced to a Laplacian term, whose
coefficient is a decreasing function of $\nu$, and becomes negative at
${\nu}_{\rm c}^{\phi}$.  Thus the homogeneous state is stable for $\nu
\le \nu_{\rm c}^{\phi}$, and it becomes more unstable as $\nu$
increases.

\subsection{Stripe state}

The behavior below the stability border is investigated numerically.
The forced CGLE in one dimension is integrated using a pseudospectral
method for various $\nu$ and $B$ \cite{f91}.  The spatial resolution
$\Delta x$ and time step $\Delta t$ used are $0.1$ and $0.01$,
respectively.  Also, a periodic boundary condition is used.  For most
cases, the linear size of the system is chosen to be $4096$, and the
time interval of $2 \times 10^{4}$ is used.  Larger systems for longer
intervals are also studied, and no change in the behavior is observed.

The numerical integrations confirm the prediction that the homogeneous
state is stable above the stability border.  It is found that the
state undergoes a supercritical bifurcation to a spatially periodic
static ``stripe'' state as $B$ decreases below the border, and the
modulation amplitude of the state, defined as $\delta \phi = \sqrt
{\langle (\phi - \langle \phi \rangle_{x})^2 \rangle_{x}}$,
behaves as $\sqrt{B_{\rm s} - B}$ close to the border
[Fig.~\ref{fig:stripe}(a)].  The wave number of the stripe state is
found to agree very well with $k_{\rm max}$ of Eq.~(\ref{Eq:kmax}),
especially near the border.  In order to check how the nature of 
the transition depends on the unforced dynamics, the transition from
a homogeneous state is examined for four different values of
$(\alpha,\beta)$: $(-2,2)$, $(-1.11, 1)$, $(-2,0)$, and 
$(-0.75,0.5)$.  It is found that a supercritical transition to the
stripe state is observed for the first two cases belonging to the
Benjamin-Feir (BF) unstable region, while a transition to a disordered
structure is observed for the other two cases belonging 
to the BF stable region. 

As $B$ decreases further, the stripe state becomes unstable to a
fluctuating stripe or ``kink-breeding'' state, depending on $\nu$
\cite{Chate3}.  The stability border of the stripe state is
determined, and is plotted in Fig.~\ref{fig:phase-d}(a).  Again, the
border is not symmetric to the $\nu = \alpha$ line.  The region of the
stripe state is much broader on the $\nu > \alpha$ side.  Moreover, it
extends to the region where a locked homogeneous state does not exist.
The origin of the asymmetry will be discussed using a phase equation,
and is found to be very different from the case of the homogeneous
state.

\section{Phase Equation}

\subsection{Derivation}

An approximate phase equation can be derived from Eq.~(\ref{Eq:FCGLE})
as follows.  Define small variables $r = R - R_{0}$ and $\phi = \Phi -
\Phi_{0}$, and assume that the time scale for $\phi$ is much larger
than that for $r$.  The variable $r$ is then slaved to $\phi$.
Starting from Eq.~(\ref{Eq:RP}), it can be shown that
\begin{equation}
\label{Eq:rs}
r = {R_0 \over 3R_0^2-1} \left[(1-R_0^2) + {B \over R_0} \cos(\Phi_0 +
\phi) - \beta \phi_{xx} - \phi_{x}^{2} \right],
\end{equation}
where additional terms higher than the second order in $\phi$ are
ignored, and $B$ is assumed to be small.  Substituting this to the
phase part of Eq.~(\ref{Eq:RP}),
\begin{eqnarray}
\label{Eq:Phase}
R_0 \partial_{t} \phi &=& \nu R_0 - \alpha R_0^3 - B \sqrt{1+a^2}
\sin(\Phi_0 +\phi+\delta) \nonumber\\ && + b \phi_{x}^2 + c \phi_{xx} + d
\phi_{xxxx} + e,
\end{eqnarray}
where $a, b, c, d, e, \delta$ are constants depending on $\alpha,
\beta, \nu$, and $R_0$.  Since $R_0$ is $1$ at $\nu = \alpha$, and is a
slowly varying function of $\nu$, it is expected that setting $R_0 =
1$ does not change the qualitative behavior of the equation.  On the
other hand, the constants are simplified to
\begin{eqnarray}
\label{const}
a &=& \alpha - (\nu - \alpha) / 2, \nonumber \\
b &=& \alpha - \beta - (\nu - \alpha)/2, \nonumber \\
c &=& 1 + \alpha \beta - \beta(\nu-\alpha)/2, \nonumber \\
d &=& -\beta^2 / 2, \nonumber \\
e &=& 0, \nonumber \\
\delta &=& \tan^{-1} (a).
\end{eqnarray}
For the remainder of the paper, $R_0$ will be set to $1$ in the
equation.  Note that Eq.~(\ref{Eq:Phase}) is a generalized version of
the phase equation obtained by Coullet and Emilsson, which is derived
for the special case of $\nu \simeq \alpha$ \cite{Coullet}.  Also,
$\phi_{xxxx}$ term is added for the stability of the solution in
the phase and defect turbulence regions.

\subsection{Homogeneous state}

\begin{figure}[ht]
\centerline{\psfig{figure=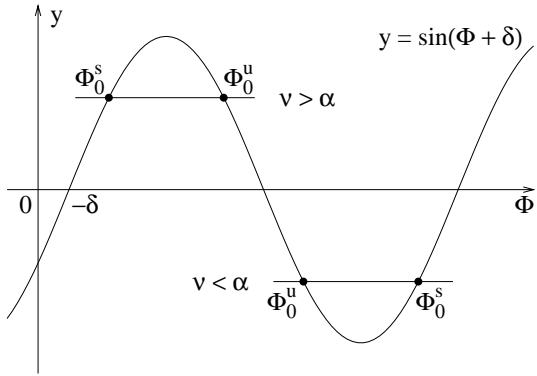,angle=270,width=0.45\textwidth}}
\caption{A schematic view of the determination of the homogeneous
solution of the phase equation Eq.~(\ref{Eq:Phase}): There exist one
stable $\Phi_0^{\rm s}$ and one unstable $\Phi_0^{\rm u}$ solutions.
Here, the one with positive $\cos(\Phi_0+\delta)$ is stable.  For $\nu
> \alpha, \Phi_0^{\rm u} > \Phi_0^{\rm s}$, while $\Phi_0^{\rm u} <
\Phi_0^{\rm s}$ for $\nu < \alpha$.}
\label{fig:flat-sol}
\end{figure}

The phase equation is studied in a way parallel to the analysis of the
forced CGLE.  Homogeneous states, given by
\begin{equation}
\label{Eq:P0}
\Phi_0 = \sin^{-1} ({\nu - \alpha \over B \sqrt{1+a^2}}) - \delta
\end{equation} 
exist for $B \ge (\nu - \alpha) / \sqrt{1+a^2}$.  There exist two
solutions $\Phi_0^{\rm s}$ and $\Phi_0^{\rm u}$, in the $[0, 2\pi]$
interval satisfying Eq.~(\ref{Eq:P0}) as shown in
Fig.~\ref{fig:flat-sol}.  The $\Phi_0^{\rm s}$ solution is stable
under homogeneous perturbation, while the $\Phi_0^{\rm u}$ solution is
unstable.  A linear stability analysis of the stable homogeneous state
shows that the maximum growth rate is
\begin{equation}
\label{Eq:rate}
\lambda_{\rm max}^{\phi} = - B \sqrt{1+a^2} \cos(\Phi_0+\delta) -
c^2/4d
\end{equation}
for the mode with wave number $k_{\rm max}^{\phi} = \sqrt{c/2d}$ (if $c
\le 0$).  This state is found to be linearly stable above the stability
border $B_{\rm s}^{\phi}$, which is given as
\begin{equation}
\label{Eq:Bs}
B_{\rm s}^{\phi} = \sqrt{
{(c^2/4d)^2 + (\nu-\alpha)^2 \over 1+a^2}}.
\end{equation}
The existence and stability borders are plotted in
Fig.~\ref{fig:phase-d}(b) for $\alpha = -3/4, \beta = 2$.  Note that
the shapes of the borders are qualitatively the same as those of the
forced CGLE: the stability border is asymmetric to the $\nu = \alpha$
line, and the two borders meet for $\nu \le \nu_{\rm c}^{\phi}$.
Since the wave number of the most unstable mode should be zero at $\nu
= \nu_{\rm c}^{\phi}$, one arrives at
\begin{equation}
\nu_{\rm c}^{\phi} = {2 + 3 \alpha \beta \over \beta}. 
\end{equation}
For the above values of $\alpha$ and $\beta$, $\nu_{\rm c}^{\phi} =
-5/4$, which is comparable to the value of $\nu_{\rm c}$ for the
forced CGLE.

The simple structure of the phase equation makes its interpretation
simple.  The reason for the instability is that $c$ can be negative,
while the $\phi_{xxxx}$ term always tries to suppress such an
instability.  The value of $c$ remains positive for $\nu < \nu_{\rm
c}^{\phi}$, and the homogeneous state is stable.  As $\nu$ increases
further, $c$ becomes negative.  Since $c$ is a decreasing function of
$\nu$, the instability becomes stronger with increasing $\nu$, which
explains the fact that the difference between the existence and
stability border increases with $\nu$.

\subsection{Stripe state}

\begin{figure}[ht]
\centerline{\psfig{figure=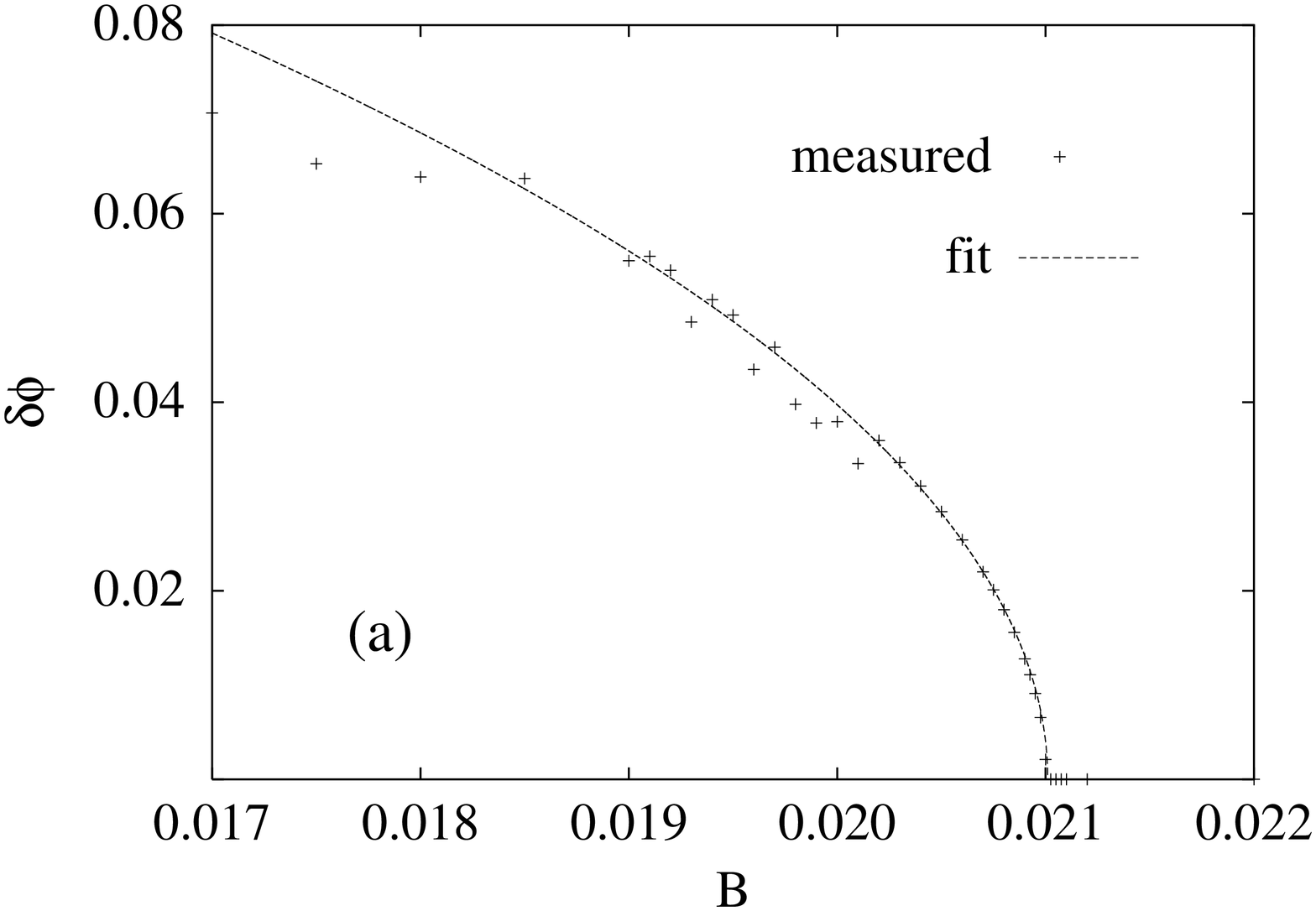,angle=270,width=0.45\textwidth}}
\centerline{\psfig{figure=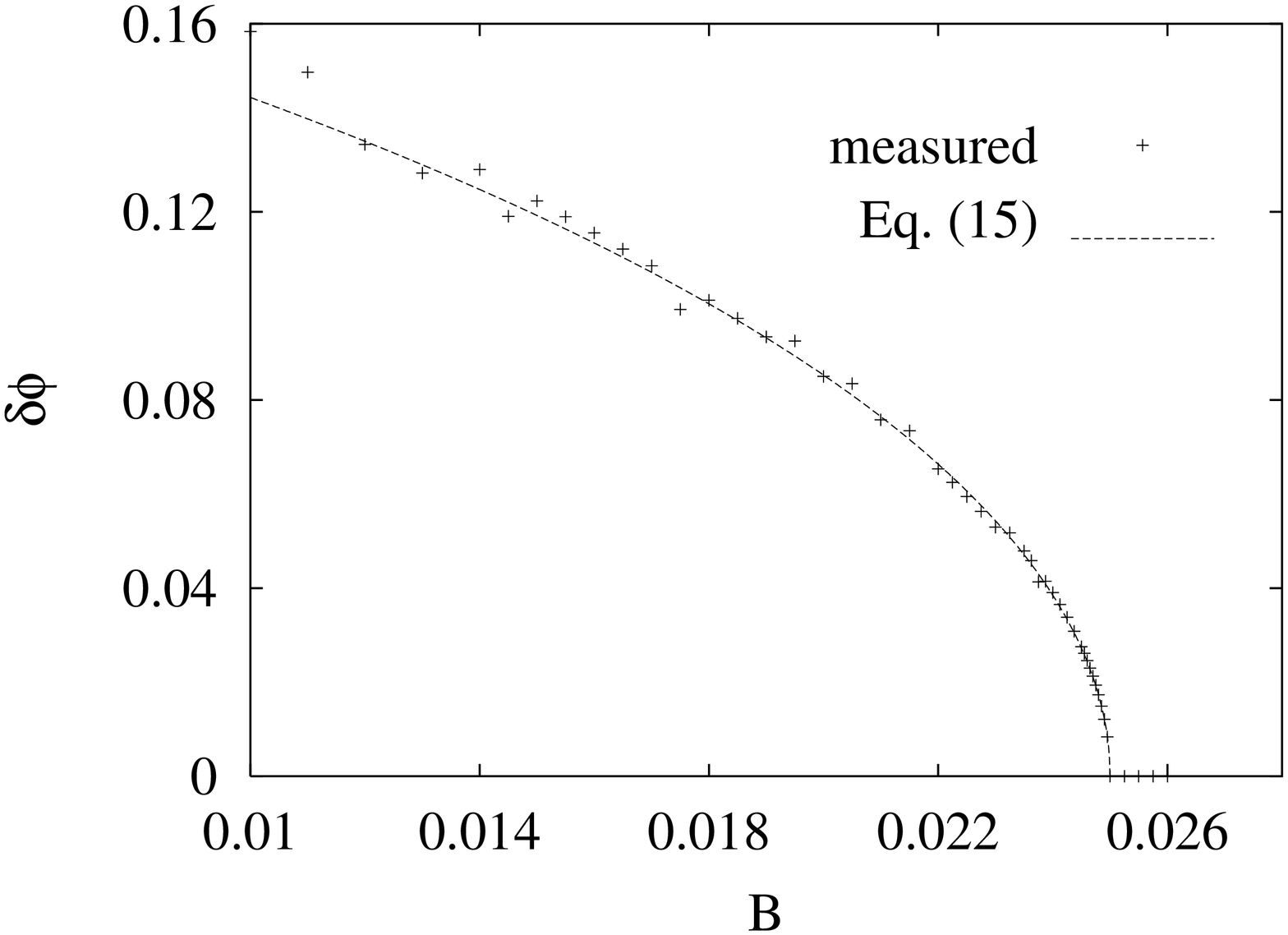,angle=270,width=0.46\textwidth}}
\caption{(a) Modulation amplitude $\delta \phi$ of the stripe
solution of the forced CGLE Eq.~(\ref{Eq:FCGLE}) is shown against $B$
for $\nu = -0.75$.  Also shown is a square root fit $D \sqrt{B_{\rm s}
- B}$ with $D = 1.25$ and $B_{\rm s} = 0.02101$.  (b) $\delta \phi$ of
the stripe solution of the phase equation Eq.~(\ref{Eq:Phase}) is
shown for $\nu = -0.75$.  Analytic expression Eq.~(\ref{Eq:series})
with the coefficients given by Eq.~(\ref{Eq:stsol}) is found to be a
good approximation.}
\label{fig:stripe}
\end{figure}

The behavior of the phase equation below the stability border is
studied numerically.  As one crosses the border, the homogeneous state
goes through a supercritical bifurcation to a stripe state, and the
modulation amplitude $\delta \phi = \sqrt {\langle (\phi - \langle
\phi \rangle_{x})^2 \rangle_{x}}$ behaves as $\sqrt{B_{\rm
s}^{\phi}-B}$ close to the border.  A typical dependence of $\delta
\phi$ on $B$ is shown in Fig.~\ref{fig:stripe}(b), where $\nu =
-0.75$.  As $B$ decreases further, the stripe state becomes unstable.
The stability border of the stripe state determined numerically is
plotted in Fig.~\ref{fig:phase-d}(b).  Again, the border is not
symmetric to the $\nu = \alpha$ line, and even extends below the
existence border of the homogeneous state.  Although the phase
equation is simpler than the forced CGLE, their qualitative behaviors
are essentially the same, at least for the homogeneous and stripe
states.


The simple structure of the phase equation allows an analytic
expression for the stripe solution.  The solution may be expanded in
terms of harmonic functions
\begin{eqnarray}
\label{Eq:series}
\phi (x) = \phi_1 && + S_1 \sin{(k_0x)} + C_1 \cos{(k_0x)}
\nonumber\\ && + S_2 \sin{(2k_0x)} + C_2 \cos{(2k_0x)},
\end{eqnarray}
where higher harmonics are ignored.  $k_0$ is the wave number of the
most unstable mode, and the coefficient $C_1$ can always be set to $0$
by choosing an appropriate origin.  Substituting it in
Eq.~(\ref{Eq:Phase}), we find
\begin{eqnarray}
\label{Eq:A1B2}
&& S_2 = 0, \nonumber \\
\sin(\Phi_0\!+\!\phi_1\!+\!\delta)~&& =
{\nu - \alpha + (S_1^2+4C_2^2) ~bc/4d \over B \sqrt{1+a^2} ~
(1-(S_1^2+C_2^2)/4)}, \nonumber\\
S_1^2 = 4 C_2 &&~{-B \sqrt{1+a^2} \cos(\Phi_0\!+\!\phi_1\!+\!\delta) + 2c^2/d \over
B \sqrt{1+a^2} \sin(\Phi_0\!+\!\phi_1\!+\!\delta) - bc/d}, \nonumber\\
C_2 = 2 &&~{-B\sqrt{1+a^2}\cos(\Phi_0\!+\!\phi_1\!+\!\delta) - c^2/4d \over
B \sqrt{1+a^2} \sin(\Phi_0\!+\!\phi_1\!+\!\delta) + 2bc/d},
\end{eqnarray}
which can be solved numerically for $S_1, C_2$ and $\phi_1$.  Near the
stability border $B_{\rm s}^{\phi}$, an approximate analytic solution
can be obtained, which is
\begin{eqnarray}
\label{Eq:stsol}
S_1 & \propto & \sqrt{B_{\rm s}^{\phi} - B}, \nonumber \\ C_2 &
\propto & S_1^2, \nonumber \\ \phi_1 & \simeq & {1 \over 4 B
\sqrt{1+a^2} \cos (\Phi_0 + \delta)} ~(\nu - \alpha + {bc \over d})
~S_1^2,
\end{eqnarray}
where the proportionality constants are rather complex except for the
case of $\phi_1$.  The analytic solution agrees well with the results
using numerical integration: as shown in Fig.~\ref{fig:stripe}(b), the
modulation amplitude $\delta \phi$ vs $B$ curve obtained from the
above expression is in good agreement with the corresponding numerical
values.

\begin{figure}[ht]
\centerline{\psfig{figure=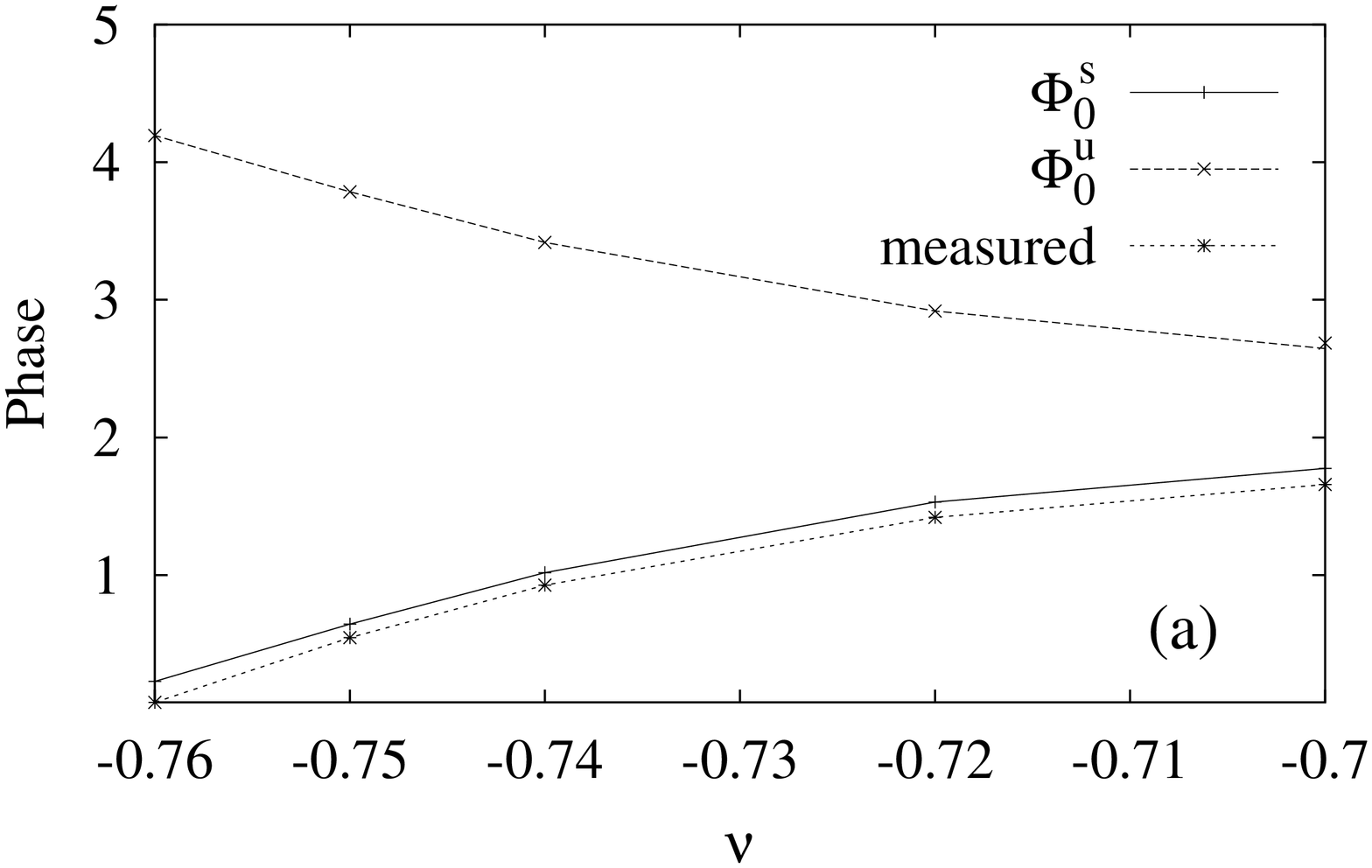,angle=270,width=0.45\textwidth}}
\centerline{\psfig{figure=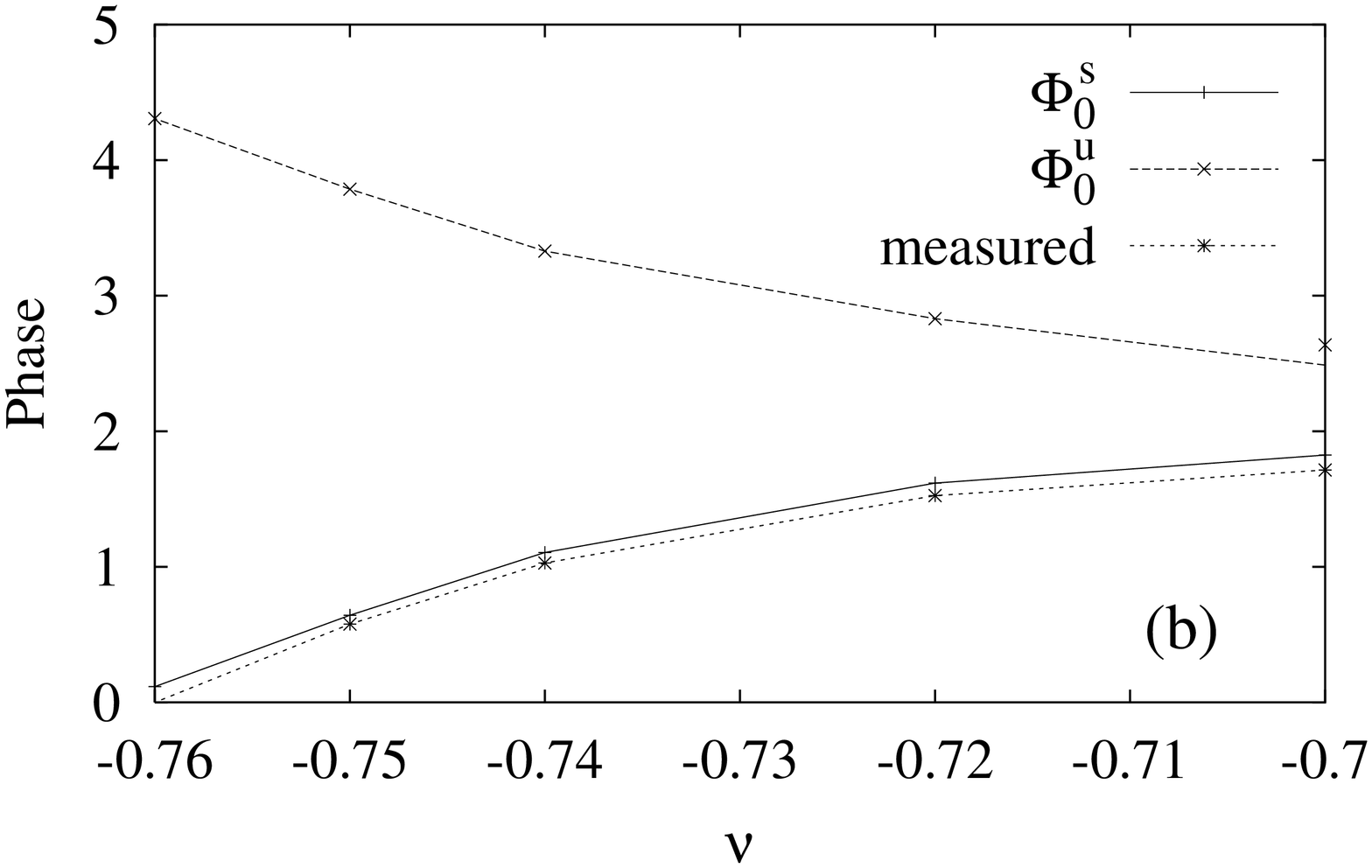,angle=270,width=0.45\textwidth}}
\caption{(a) The stable $\Phi_0^{\rm s}$ and unstable $\Phi_0^{\rm u}$
homogeneous solutions of the phase equation Eq.~(\ref{Eq:Phase}) just
below the stability border ($B - B_s^{\phi} = 5 \times 10^{-3}$) are
plotted against $\nu$.  The average phase of the stripe solution of
the equation for the same $B$ is also plotted.  The average phase is
smaller than $\Phi_0^{\rm s}$, which moves away from (toward)
$\Phi_0^{\rm s}$ for $\nu > \alpha$ ($\nu < \alpha$).  (b) The same
plot for the forced CGLE.  The qualitative behavior is identical.}
\label{fig:shift}
\end{figure}

The analytic solution confirms not only the square root dependence of
$A_1$, but it also provides an explanation for the asymmetry of the
stability border of the stripe solution.  Shown in
Fig.~\ref{fig:shift}(a) are two homogeneous solutions---stable
$\Phi_0^{\rm s}$ and unstable $\Phi_0^{\rm u}$---of the phase equation
with $B$ a little below the stability border.  As $B$ decreases from
the homogeneous toward the stripe region, the modulation amplitude
around $\Phi_{0}^{\rm s}$ increases with decreasing $B$.  For
sufficiently small $B$, $\phi(x)$ at certain $x$ approaches the
unstable fixed point $\Phi_0^{\rm u}$, which then makes the stripe
solution unstable.  Note that $\phi_1$ in Eq.~(\ref{Eq:stsol}) is
nonzero---it is negative when $\nu$ is not very different from
$\alpha$.  Thus, the average phase of a stripe state is shifted toward
a value smaller than $\Phi_0^{\rm s}$.  The average phase of the
stripe solution measured using numerical integration is also shown,
which confirms the shift.  As shown in Fig.~\ref{fig:flat-sol}, the
shift moves the stripe solution toward (away from) $\Phi_0^{\rm u}$
for $\nu < \alpha$ ($\nu > \alpha$), resulting in the asymmetry (a
related argument is given in \cite{Park}).  The situation is entirely
similar for the forced CGLE.  The phase of the stable and unstable
solution is plotted against $\nu$ in Fig.~\ref{fig:shift}(b)
\cite{note}.  Also plotted is the average phase of the stripe
solution.  Here again, the average phase is shifted toward (away from)
the unstable solution for $\nu < \alpha$ ($\nu > \alpha$).


\section{Conclusion}

Despite its simplicity, the forced CGLE displays a large variety of
phenomena.  The homogeneous and stripe states are mainly discussed
here, and the phase equation is found to be very useful in
understanding the stability borders of the forced CGLE.  For a
sufficiently large forcing amplitude, a homogeneous state with no
spatial structure is observed.  The state becomes unstable to a
spatially periodic stripe state via a supercritical bifurcation as the
forcing amplitude decreases. We obtained an analytic solution for the
stripe state of the phase equation, through which an argument for the
asymmetry of the stability border of the state is formulated.  As $B$
decreases further, more complex behaviors such as, the kink breeding,
``depinning,'' and ``roughening,'' are expected, which are currently
under investigation.


This work was supported in part by grant No. 2000-2-11200-002-3 from
the BRP program of the KOSEF.  J. Lee was supported by Creative
Research Initiatives of the Korean Ministry of Science and Technology.
The computation was done on a PC cluster in the Supercomputing Center
of Korea Institute of Science and Technology Information.  We acknowledge
for the generous computer time, and thank Mr. Jeong-Woo Hong for
helpful technical assistance.




\end{multicols}{2}

\begin{thebibliography}{99}
\bibitem{note} For the forced CGLE, there exist three homogeneous
solutions, instead of two for the phase equation.  The third solution,
which belongs to the small $|A|$ branch, is irrelevant in the
discussion of the stripe solution.
\end{thebibliography}
\end{document}